\documentclass[journal=jctcce,manuscript=article,layout=twocolumn]{achemso}

\usepackage{graphicx}
\usepackage{dcolumn}
\usepackage{bm}
\usepackage{hyperref}
\hypersetup{colorlinks,allcolors=blue}
\usepackage{ulem}
\usepackage{xcolor}
\usepackage{color,soul}

\DeclareUnicodeCharacter{0308}{\"{u}}

\title{Efficient parameterization of transferable Atomic Cluster Expansion for water}

\author{Eslam Ibrahim}
\email{eslam.saadibrahim@rub.de}
\author{Yury Lysogorskiy}
\email{yury.lysogorskiy@rub.de}
\author{Ralf Drautz}
\email{ralf.drautz@rub.de}
\affiliation{ICAMS, Ruhr Universit\"at Bochum, 44780 Bochum, Germany}

\begin{document}




\begin{abstract}
We present a highly accurate and transferable parameterization of water using the atomic cluster expansion (ACE).
To efficiently sample liquid water, we propose a novel approach that involves sampling static calculations of various ice phases and utilizing the active learning (AL) feature of ACE-based D-optimality algorithm to select relevant liquid water configurations, bypassing computationally intensive ab-initio molecular dynamics (AIMD) simulations.
Our results demonstrate that the ACE descriptors enable a potential initially-fitted solely on ice structures which is later upfitted with few configurations of liquid, identified with active learning to provide an excellent description of liquid water. 
The developed potential exhibits remarkable agreement with first-principles reference, accurately capturing various properties of liquid water, including structural characteristics such as pair correlation functions, covalent bonding profiles, and hydrogen bonding profiles, as well as dynamic properties like the vibrational density of states, diffusion coefficient and thermodynamic properties such as the melting point of the ice Ih.
Our research introduces a  new and efficient sampling technique for machine learning potentials in water simulations, while also presenting a transferable interatomic potential for water that reveals the accuracy of first principles reference. This advancement not only enhances our understanding the relationship between ice and liquid water at the atomic level, but also opens up new avenues for studying complex aqueous systems.
\end{abstract}

\maketitle
 
\section{Introduction}
Understanding and predicting the properties of water is crucial for numerous fields, including chemistry, materials science, and biology~\cite{jeffrey1997introduction,chaplin2019structure}. 
Despite the simplicity of the H$_2$O molecule, the simulation of molecular-level dynamics of water and its phases~\cite{monserrat2020liquid,cheng2021phase,reinhardt2022thermodynamics,gartner2022liquid} remains challenging due to the presence of strong intermolecular interactions~\cite{bernal1933theory,debenedetti2003supercooled, coulson1966interactions, batista1998molecular, silvestrelli1999water, badyal2000electron}. 
The choice of ab-initio functionals significantly impacts simulation outcomes~\cite{gillan2016perspective, cisneros2016modeling, medders2013critical} and several investigations were carried out to evaluate predictions of water properties~\cite{laasonen1993ab,tuckerman1994ab,pestana2017ab, ruiz2018quest, gillan2016perspective, cisneros2016modeling, medders2013critical, chen2017ab, sun2015strongly, sun2016accurate,sprik1996ab}.


Ice in its various crystalline forms exhibits a wide range of structural motifs, each characterized by a specific arrangement of water molecules.
Liquid water, on the other hand, lacks long-range order but forms transient hydrogen bonds, which leads to close similarities in structure between ice and water~\cite{errington2001relationship}. Understanding the relation between the atomic structure of different ice phases and liquid water is crucial for unraveling the intricate structure-property relationships that govern the properties of water~\cite{monserrat2020liquid}. 

Experiments such as X-ray diffraction~\cite{krack2002ab} and infrared spectroscopy~\cite{sharma2005intermolecular,zhang2011first} have provided fundamental insight into the atomic structure of ice and liquid water. Atomic scale analysis is challenging~\cite{herbert2019structure,svoboda2020real,nishitani2019binding,turi2005characterization, savolainen2014direct,gartmann2019relaxation} 
and computational approaches, particularly molecular dynamics simulations, have emerged as a powerful tool to complement experimental studies. Empirical models and interatomic potentials led the way~\cite{mackerell1998all, jorgensen1983comparison, price2004modified, abascal2005general}. 

In recent years, machine learned interatomic potentials have enabled much more accurate predictions ~\cite{bartok2017machine, kulik2022roadmap, deringer2021gaussian}.
and a substantial number of machine learned interatomic potentials for liquid water were developed~\cite{cheng2019ab, lan2021simulating, daru2022coupled, liu2022toward, piaggi2021phase, kapil2024first, monserrat2020liquid, reinhardt2021quantum, cheng2021phase, reinhardt2022thermodynamics, schran2021machine}.


We present a highly accurate and transferable parameterization for ACE that we demonstrate captures 
the intricacies of water atomic interactions with the accuracy of the first-principles reference. Different from most other MLPs, we do not base our parameterization on data obtained from ab initio molecular dynamics simulations in water. Instead, we exploit the fact that the local atomic motifs present in liquid water can also be obtained from ice with displaced atomic positions, which provides a more efficient parameterization strategy. 


The remainder of this article is organized as follows: Section~\ref{sec:methods} provides an overview of the computational methods employed in this study.
Section~\ref{sec:results} focuses on the relation between ice phases and liquid water together with a various validation tests.  
Finally, Section~\ref{sec:conc} summarizes the key results and discusses their implications for future studies in the field of water science.

\section{Methods}
\label{sec:methods}

\begin{figure*}[hbt!]
    \centering
    \includegraphics[width=15cm]{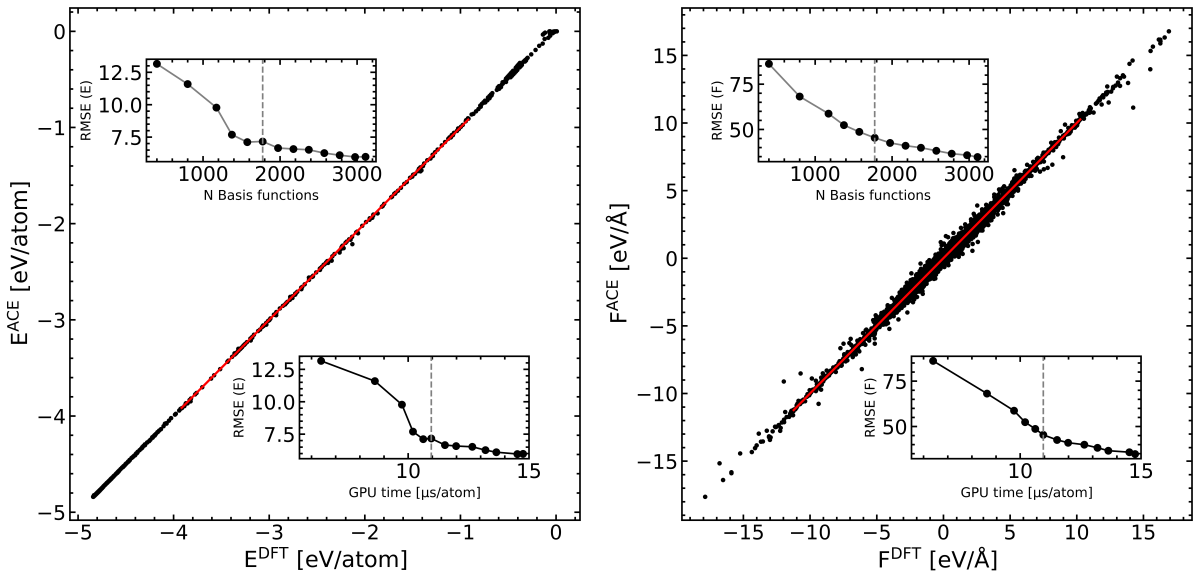}
    \caption{Comparison of ACE predictions for energies and forces to DFT reference. Insets show the convergence of RMSE with respect to the number of ACE basis functions and the GPU time in units of eV/atom and eV/\AA, respectively for NVIDIA GeForce RTX 3060 Ti (compute architecture 8.6).}
    \label{fig:train_data}
\end{figure*}

\subsection{Reference calculations}
We utilized the Vienna Ab initio Simulation Package (VASP)~\cite{kresse1994google,kresse1996efficiency,kresse1996efficient} for conducting density functional theory (DFT) calculations. All DFT calculations were performed employing the Perdew-Burke-Ernzerhof (PBE) exchange-correlation functional~\cite{perdew1996generalized}.
To account for van der Waals (vDW) dispersion interactions, we incorporated the Grimme D3 method~\cite{grimme2010consistent,grimme2011effect}, where earlier studies had shown that it benefits from intrinsic functional error cancellations~\cite{pestana2017ab}, leading to a description of water that compares very well with experiment in simulations with classical nulei.A k-mesh density of 0.125/\AA, a plane wave energy cutoff of 450\,eV.

The initial set of 83\, ice structures were fully relaxed, for all other configurations that were generated using active learning energies and forces were calculated solely in  static calculations.

\subsection{Molecular dynamics simulations}
The Large-scale Atomic/Molecular Massively Parallel Simulator (LAMMPS)~\cite{LAMMPS} with the ML-PACE pair style~\cite{lysogorskiy2021performant} were employed to carry out molecular dynamics (MD) simulations. The vibrational density of states (VDOS) were obtained from the velocity auto-correlation function (VACF) as implemented in i-pi~\cite{kapil2019pi}.

\subsection{Atomic cluster expansion}
ACE~\cite{drautz2019atomic,dusson2022atomic} training was carried out using the package Pacemaker~\cite{bochkarev2022efficient}. We employed a Finnis-Sinclair-type mildly non-linear representation of the atomic energy that incorporates two atomic properties, which are represented by linear ACE basis expansions~\cite{drautz2019atomic,lysogorskiy2021performant,bochkarev2022efficient}. The non-linear representation is motivated by the second-moment approximation and has been demonstrated to be efficient for metals~\cite{bochkarev2022efficient,lysogorskiy2021performant,e_ibrahim_magnesium} as well as covalently bonded materials~\cite{qamar2023atomic,bochkarev2022multilayer}, for non-collinear magnetic degrees of freedom \cite{rinaldi2024non}, message passing~\cite{bochkarev2023atomic} and ACE-based graph basis functions ~\cite{bochkarev2024graph}.

\subsection{Training dataset}

\subsubsection{Generation of diverse ice phases}
We started by generating a diverse set of structures using the genice code~\cite{Matsumoto:2017bk, Matsumoto:2021}, which implements a graph-theory based algorithm to ensure topological characteristics of proton-ordered as well as proton-disordered ice. We selected intially 83\,ice structures with up to 310\,atoms for training a first ACE,  where we adopt the same naming convention from the genice code.
Using this ACE, we used active learning~\cite{lysogorskiy2023active} to select further ice structures with densities from 0.1\,to 17\,g/mL and with random deformations on atomic positions and cell. 
The dataset for training was obtained by constraining maximum force components to below 35\,eV/\AA\, and densities from 0.16\, to 1.82\,g/mL.


\subsubsection{Generation of liquid configurations using active learning}
To avoid computationally expensive ab-initio molecular dynamics (AIMD) simulations for sampling liquid water, we ran molecular dynamics in the isobaric-isothermal ensemble (NPT) with the ACE that was trained solely on ice structures (ACE-ice). During the simulation we kept track of the uncertainty indicator ~\cite{lysogorskiy2023active} to identify extrapolative configurations, from which we selected some for DFT calculations.


\subsection{Training procedure}

For ACE parameterization we employed a successive hierarchical basis extension with power-order ranking of basis functions~\cite{bochkarev2022efficient} up to body order five and a cutoff of  6.0\,\AA. We started training from the 83\, initial ice structures, where we include the first and last configuration from the full ionic relaxation in addition to 5\, deformed configurations for every structure. This resulted in 581\, structures with 49686\, atoms. In a first active learning step, we generated more deformed ice structures with a wider range of densities and further added O-O, H-H and O-H dimers to the training data. This dataset comprised 2432\, structures with 155068\, atoms. In the second and final active learning step, we used ACE in molecular dynamics simulations at various NPT conditions and selected liquid configurations with large extrapolation grades. This resulted in a final dataset of 2575\, structures and 173686\, atoms.
We summarize the steps in Table.~\ref{table:data}.

\begin{table}[hbt!]
\tiny
\begin{center}
  \begin{tabular}{l|ccc}
 AL step   &  Type & N configurations & N atoms\\
 \hline
 0th & Initial ices & 581 & 49686 \\
 1st & Highly deformed ices  &  2335   &  157174 \\
 2nd & O-O, O-H and H-H dimers & 2432 & 155068 \\
 3rd & AL-selected liquid water & 2575 & 173686\\
  \end{tabular}
\end{center}
\caption{Different active learning steps for dataset generation.}
\label{table:data}
\end{table}

Energies and forces  were weighted in the loss function using an energy based weighting scheme~\cite{bochkarev2022efficient} that gave higher weights to low energy structures. We truncated the expansion at 1774\,basis functions with 4844\,parameters. This resulted in a parameterization of mean-absolute-error (MAE) for energies of 2.5\, meV/atom and MAE for forces of 16.7\,meV/\AA$^{-1}$. Fig.~\ref{fig:train_data} shows that ACE reproduces the DFT reference data with excellent accuracy and illustrates that we truncated the expansion as a compromise between accuracy and computational cost.

\subsection{ACE-map}

We utilize the ACE basis functions directly to examine the local atomic environments in ice and liquid water. We employ t-distributed Stochastic Neighbor Embedding (t-SNE) ~\cite{ljpvd2008visualizing, belkina2019automated} as implemented in Scikit-learn~\cite{scikit-learn} to reduce the dimensionality of basis function space for local atomic environments of hydrogen and oxygen. The input data consisted of all basis functions for all structures from the 2nd active learning step alongside selected liquid structures at 300\,K. Our approach is similar to Ref.~\cite{monserrat2020liquid}, but differently we employ the non-linear t-SNE algorithm, which helps to maintain local structure.

\section{Results}
\label{sec:results}

\subsection{ACE-map for water}

\begin{figure*}[hbt!]
    \centering
    \includegraphics[width=15cm]{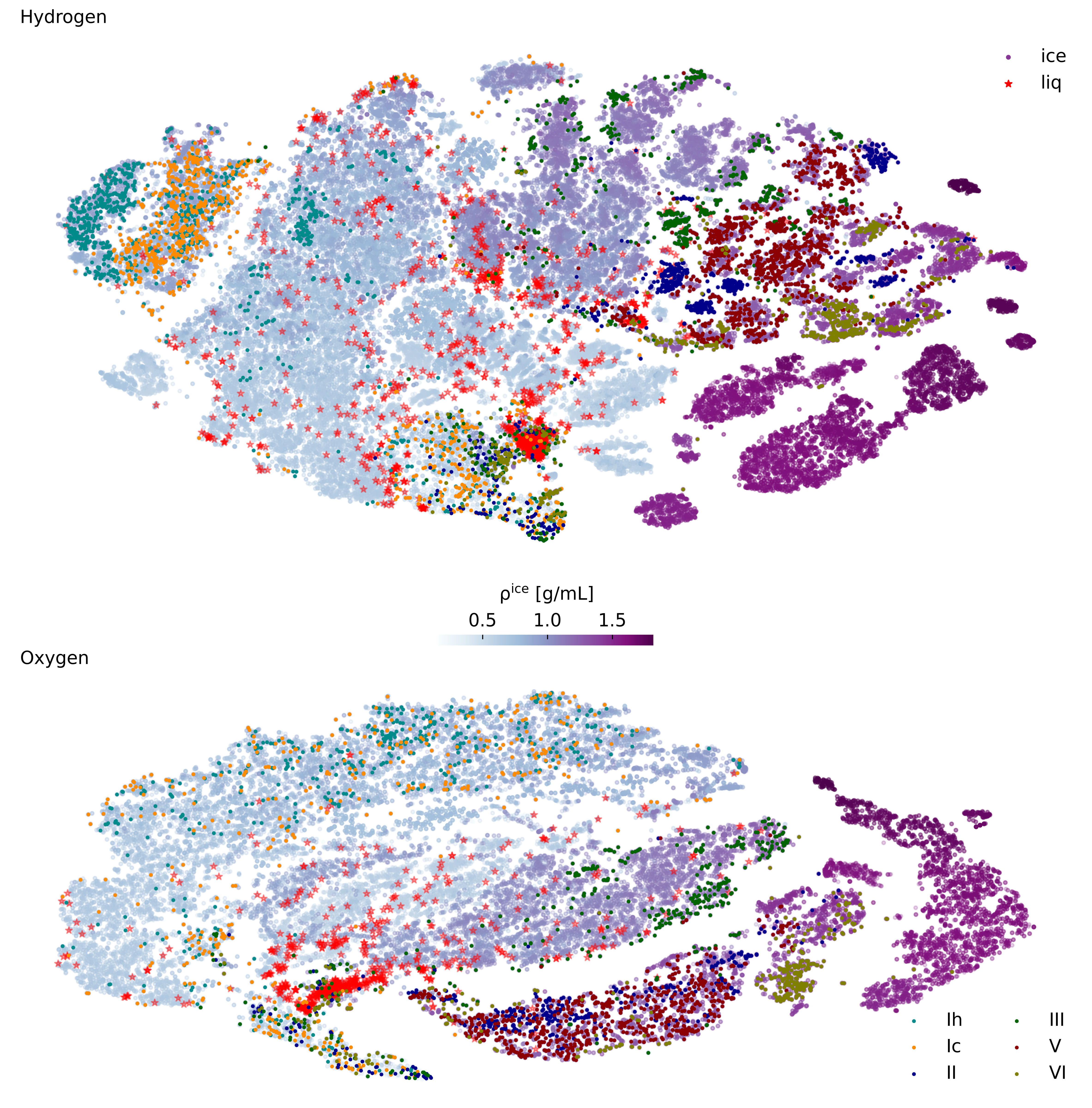} 
    \caption{T-SNE maps of ice phases and water configurations at 300 K. Hydrogen and oxygen environments are characterized separately (top and bottom, respectively). Red stars indicate environments in liquid configurations, while dots of different colors mark different ice phases, including atomic displacements in these phases. Light to bold color indicate increasing density.}
    \label{fig:acemap}
\end{figure*}

Solid and liquid atomic environments overlap in ACE-map Fig.~\ref{fig:acemap}. This coincides with the observation from Ref.~\cite{monserrat2020liquid} that liquid water contains the building blocks of ice. Here we turn this around. Ice phases including atomic displacements contain the structural motifs of liquid water. This leads us to an efficient parameterization strategy. Instead of generating a training database for liquid water from computationally expensive AIMD that requires many time steps to de-correlate atomic configurations, we employ solid ice phases with randomly displaced atomic positions. These structures are immediately de-correlated and automatically significant for the simulation of water.

Furthermore, our selection of ice phases ensures dense coverage of the PES, enabling our potential to avoid falling into sparsely sampled regions of the trained PES. This approach also facilitates stable molecular dynamics simulations of liquid water using ACE-ice without prior fitting. 
Consequently, we efficiently identify crucial liquid configurations based on extrapolation~\cite{lysogorskiy2023active}.

\subsection{Energy-volume curves for ice phases}

\begin{figure*}[hbt!]
    \centering
    \includegraphics[width=18cm]{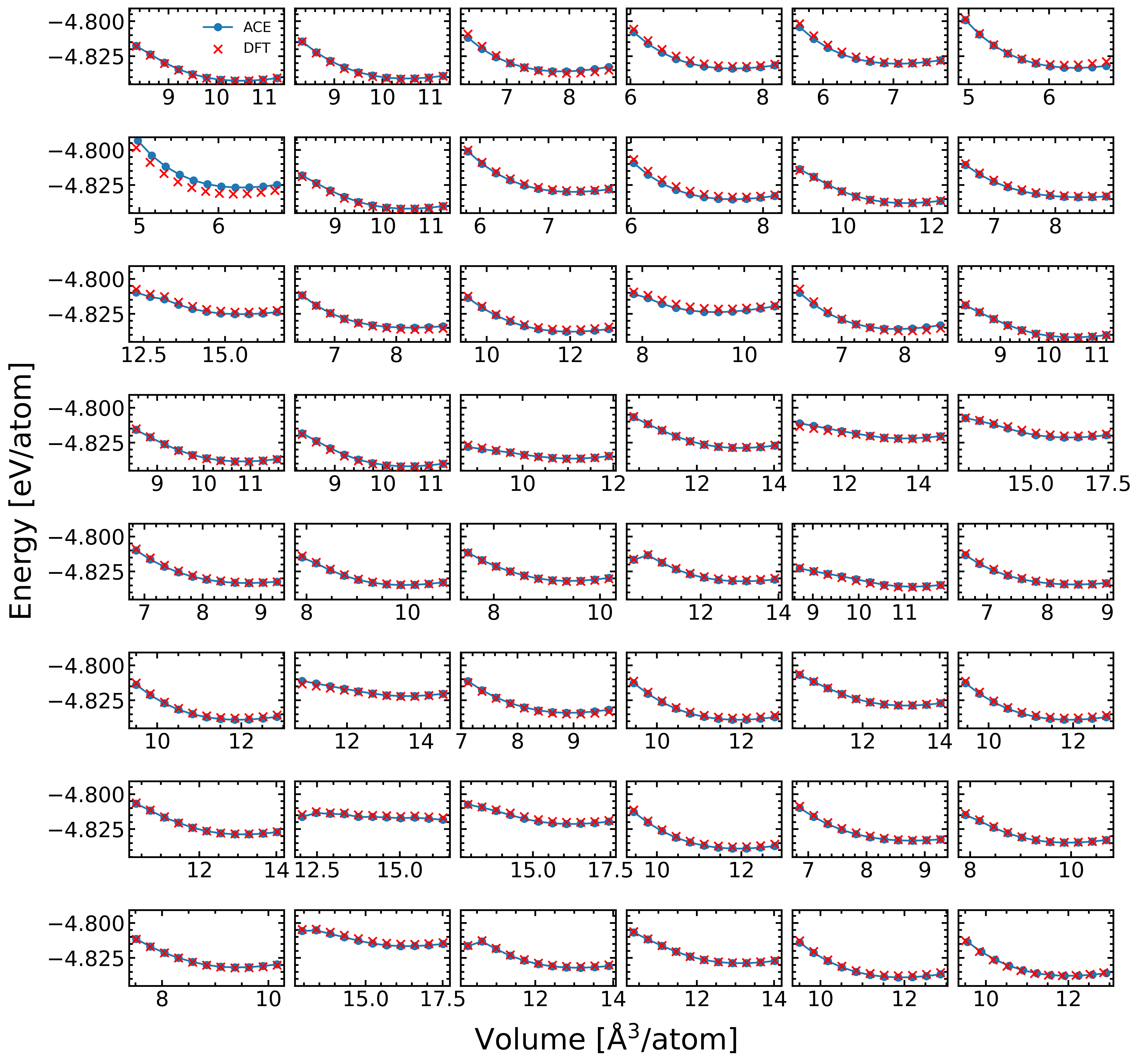}
    \caption{Energy-volume curves for various ice phases in the training dataset. The presented DFT calculations (red crosses) are not part of the training dataset.}
    \label{fig:ev_murn}
\end{figure*}

In Fig.~\ref{fig:ev_murn}, we show energy as a function of volume for a number of difference ice phases.
 The DFT calculations marked by red crosses were used for validation only. The panels demonstrate excellent transferability of ACE between a wide variety of different ice structures.


\subsection{Correlation and bonding profiles}

\begin{figure*}[hbt!]
    \centering
    \includegraphics[width=17cm]{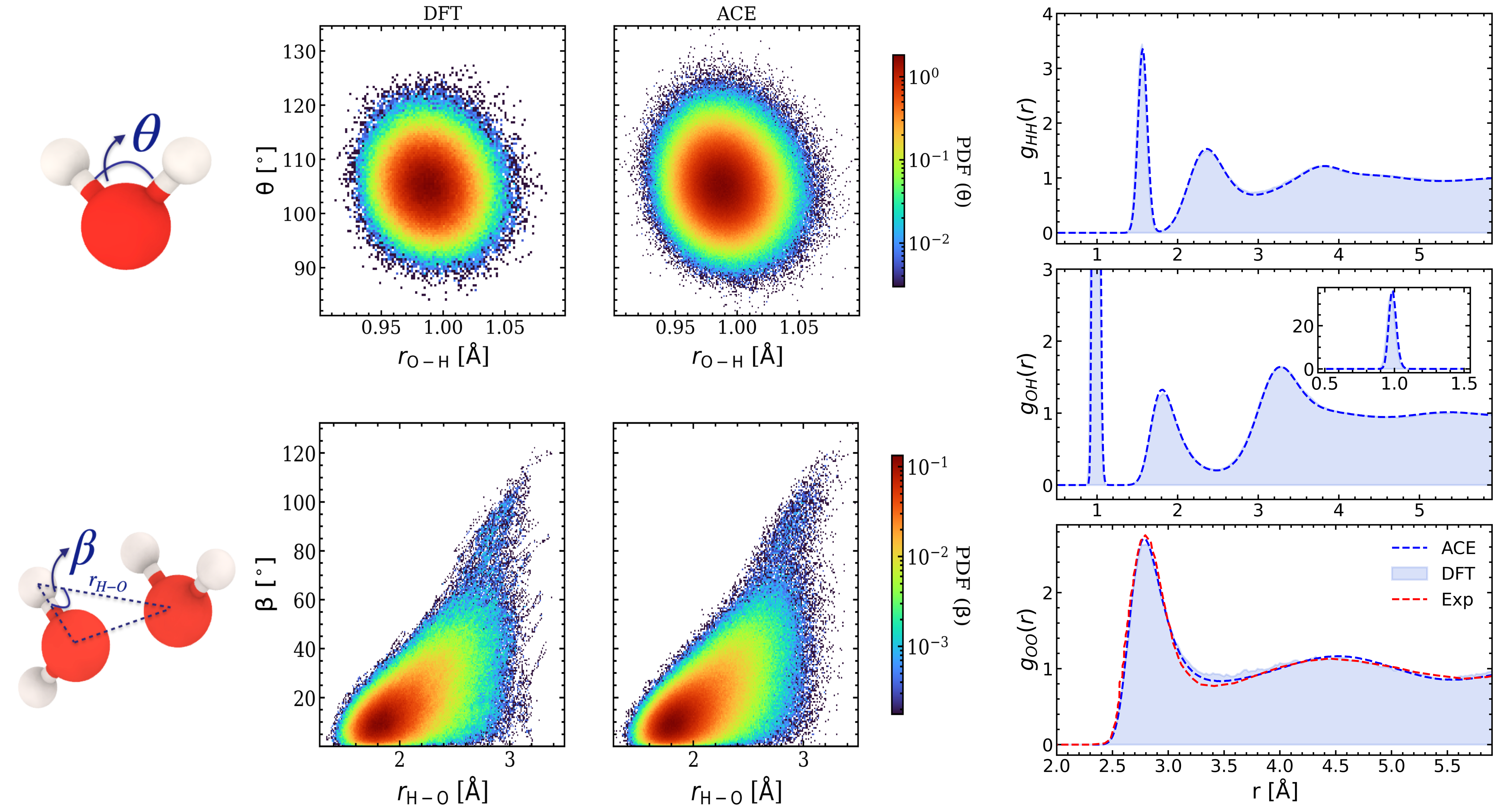}
    \caption{Radial distribution function (right) and the covalent (upper-middle) and hydrogen bonding (lower-middle) profiles in the liquid for DFT and ACE.}
    \label{fig:bonds}
\end{figure*}

We compute various pair correlation functions and profiles for different types of bonding in liquid water employing both ACE and DFT.
Specifically, we analyze the radial distribution functions (RDFs) for H-H, O-H, and O-O distributions in liquid water at a density of 0.99709\,g/mL  and temperature of 298\,K.
Our simulations utilize a cell containing 256\,molecules, simulated over 100\,ps across 7\,independent runs, totaling 700\,ps with an adapted pyiron \cite{janssen2019pyiron} implementation of neighbour lists. 
The choice of system size is guided by prior studies, indicating convergence of correlation functions with 128\,molecules~\cite{liu2022toward}.
Comparison with DFT reference data, depicted in Fig.~\ref{fig:bonds}, demonstrates excellent agreement of our ACE parameterization with the radial distribution functions from DFT. 
In particular, for validation, DFT results are obtained from simulations comprising 64 water molecules equilibrated over 1\,ps, followed by a 7\,ps production run, where none of these DFT configurations are included in our training dataset. 
While DFT results exhibit more noise compared to ACE simulations due to shorter simulation time, our AIMD data aligns with previous  findings~\cite{ruiz2018quest}.

Prior studies have highlighted the efficacy of the revPBE-D3 functional in describing O-O RDF compared to experimental results via classical MD simulations~\cite{ruiz2018quest,liu2022toward}. 
The discrepancy between quantum simulations and classical predictions is attributed to the cancellation of errors present in classical simulations but absent in quantum simulations. Notably, meta-GGA functionals offer improved descriptions in quantum simulations~\cite{ruiz2018quest,liu2022toward,marsalek2017quantum,cheng2019ab,daru2022coupled}.

We further present bonding profiles for both covalent O-H bonds and hydrogen bonds. 
Covalent O-H bonds involve hydrogen and oxygen atoms within the same molecule, while hydrogen bonds form between a hydrogen atom of one molecule and an oxygen atom of another, as illustrated in Fig.~\ref{fig:bonds}.
Our ACE parameterization consistently provides accurate predictions for both types of bonds compared to DFT.

\begin{figure}[hbt!]
    \centering
    \includegraphics[width=9cm]{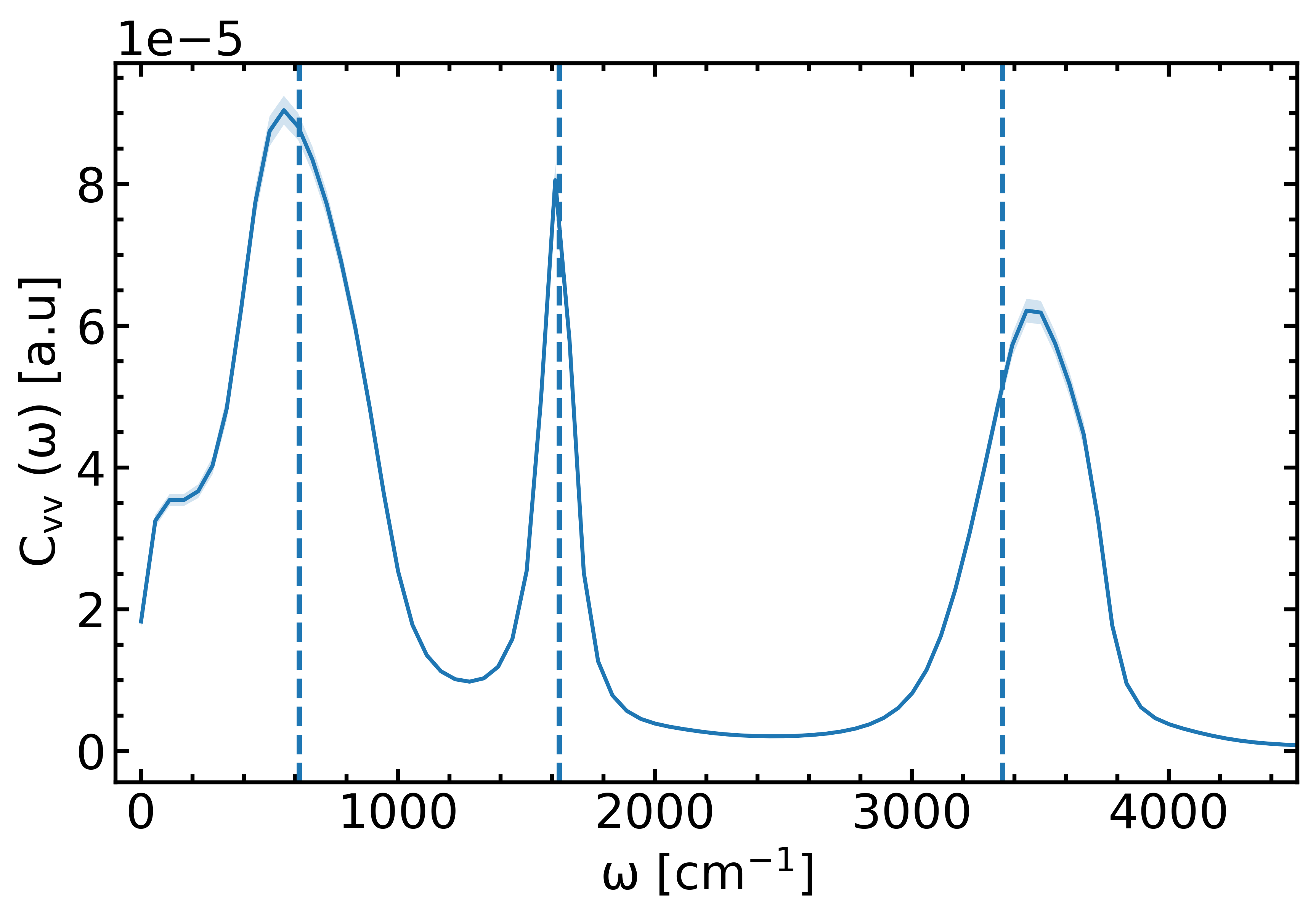}
    \caption{Fourier transform of the velocity auto-correlation function showing the VDOS of our water-ACE potential with peaks corresponding to experimental data ~\cite{liu2022toward,yao2021nuclear} highlighted by vertical lines.}
    \label{fig:vacf}
\end{figure}

\subsection{Vibrational density of states}
\label{sec:diff}

In Fig. \ref{fig:vacf}, we show the vibrational density of states for liquid water (VDOS) at 300\,K.
We calculate the VDOS from Fourier transform of the VACF $c_{vv}(\omega)$, as implemented in the i-PI engine,

\begin{equation}
    c_{vv}(\omega) = \int \langle \nu(\tau) \nu(\tau+t)\rangle_{\tau} e^{-i\omega t} dt \,.
\end{equation}

Our predictions compare well to eperimental data~\cite{yao2021nuclear} that is shown by vertical dashed lines in Fig.\ref{fig:vacf}.

\subsection{Diffusion coefficient}
We calculated the diffusion coefficient for liquid water using two different methods, from Fourier transform of VACF and the mean square displacement (MSD),

\begin{equation}
    D = \frac{1}{6} \lim_{t\to\infty} \frac{d}{dt} \langle |r(t) - r_{0}|^{2}\rangle 
\end{equation}

For MSD, we run 100\,independent NVE runs for 100 ps, totalling 10 ns, and for VACF we run 7\,independent 100\,ps, NVT~\cite{nose1984unified, hoover1985canonical} runs.
The estimated diffusion coefficient shows very good agreement with the DFT reference as well as experimental results~\cite{yao2021nuclear, pestana2017ab, marsalek2017quantum}, where both methods show close agreement, as shown in Table.~\ref{table:dc}.

\begin{table}[hbt!]
\tiny
\begin{center}
\caption{Computed and experimental diffusion coefficient.}
\label{table:dc}
  \begin{tabular}{l|cc|c}
 Method  & D [10$^{-9}$ m$^{2}$s$^{-1}$] & DFT  & Experiment~\cite{skinner2013benchmark}\\
 \hline
     VACF & 2.1  $\pm$ 0.02   & 2.22 $\pm$ 0.05~\cite{marsalek2017quantum}& \multicolumn{1}{c}{2.41 $\pm$ 0.15} \\
 MSD  & 2.1  $\pm$ 0.15   & 1.9~\cite{pestana2017ab} \\
  \end{tabular}
\end{center}
\end{table}

\subsection{Melting temperature}

\begin{figure*}[hbt!]
    \centering
    \includegraphics[width=15cm]{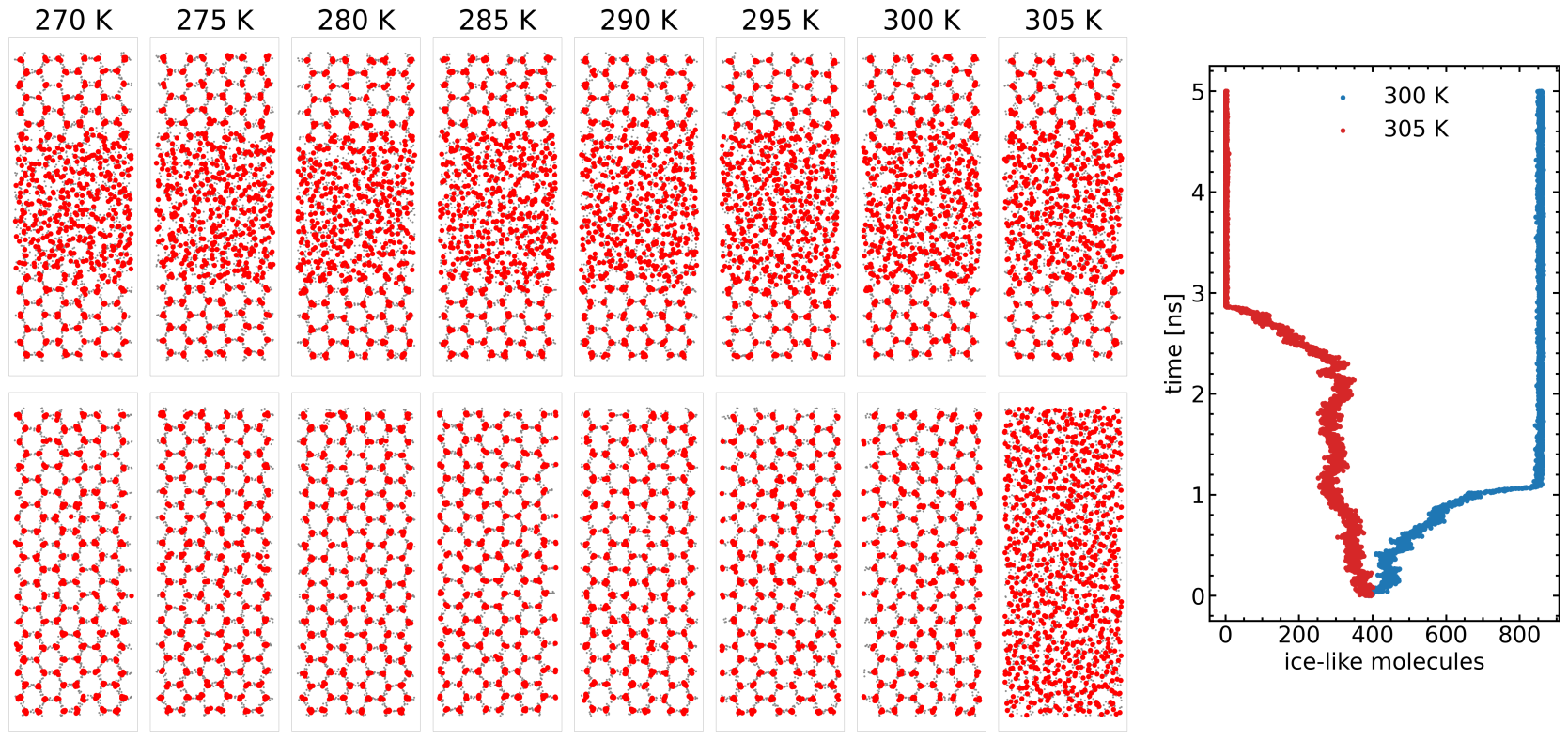}
    \caption{(left panel) Coexistence simulations of liquid water and ice Ih,  where we determine the melting point based on complete solidification or complete melting of the interface in a temperature range from 270\,K to 305\,K with an increment 5\,K. The upper row show the initial configurations and the lower row shows the final configurations after 5\,ns NPT simulation.
    (right panel) Number of ice-like-molecules at 300\,K and 305\,K, where solidification and melting take place, respectively.}
    \label{fig:tm}
\end{figure*}

We employ coexistence simulations to calculate the melting point of the ice Ih phase. We bring liquid water and ice Ih in contact in a periodic cell with 864 molecules and run NPT simulations at different temperatures in a range of 270\,K to 305\,K with steps of 5\,K at one bar, where we run each simulation for 5\,ns. After this simulation time we observe either complete melting or complete solidification. In Fig.~\ref{fig:tm}, we show the initial and  final configurations at each temperature, with complete ice growth below 305\,K and complete melting from 305\, K and above. We calculate the amount of ice-like molecules using the Auer and Frenke criterion~\cite{auer2005numerical}, implemented in pyscal~\cite{menon2019pyscal}.

\subsection{Proton recombination}

\begin{figure}[hbt!]
    \centering
    \includegraphics[width=9cm]{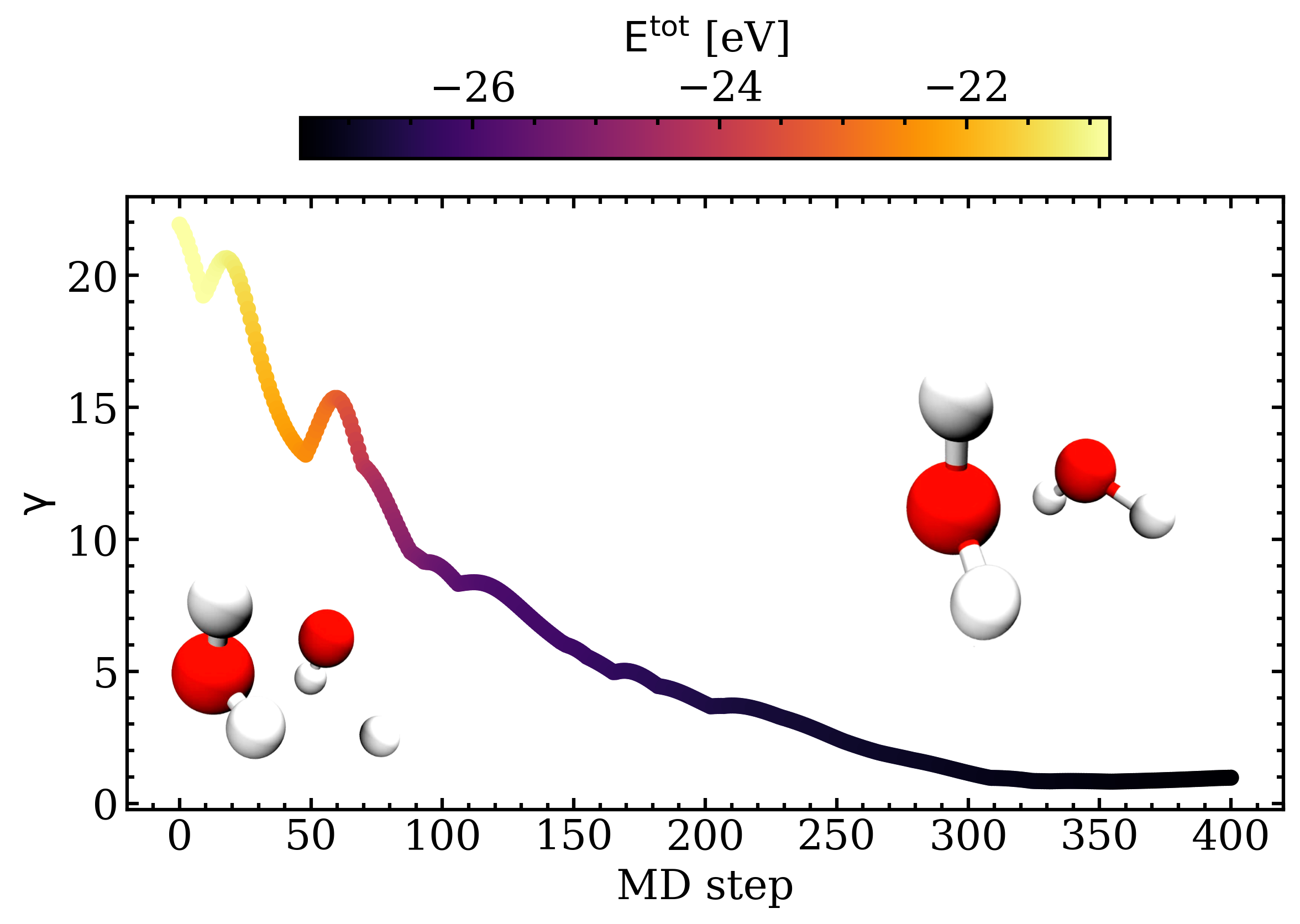}
    \caption{Simulation of proton recombination.}
    \label{fig:pr}
\end{figure}

A key feature of a transferable interatomic potential is its ability to simulate bond breaking and bond formation. Notably, our training dataset excludes configurations with broken bonds.

To test the ability of our ACE to simulate bond formation, we carried out a MD simulation with a water dimer, where one of the molecules had a single broken bond. In an NVT simulation at 300\,K we monitored both the energy and the extrapolation grade indicator.
As depicted in Fig.~\ref{fig:pr}, our ACE detected significant extrapolation during the initial MD steps, which coincided with relatively high energy states. After several hundred steps, the system stabilized with the detached proton recombining with the water molecule and the extrapolation grade decreased significantly, indicative of reliable and accurate predictions.

\section{Conclusion}
\label{sec:conc}
In this work we introduced a transferable interatomic potential for water using ACE. We followed an efficient sampling methodology for the PES that exploits the similarities of local atomic environments in liquid water and diverse ice phases. Therefore, without computationally expensive AIMD simulations, we are able to generate an ACE that shows an excellent match to first-principles computed properties and experiment.
We validate the accuracy of our ACE in different applications including energy-volume curves of diverse ice phases, structural analysis including the H-H, O-H, and O-O radial distribution functions as well as the covalent O-H and the non-directional hydrogen bonding profiles, diffusion coefficient, VDOS, melting temperature and proton recombination.

\section*{Data availability}
The initial ices DFT reference dataset (including ices coordinates) as well as potential files are made available upon acceptance of the manuscript.

\section*{Acknowledgements}
E.I acknowledges funding through the International Max Planck Research School for Sustainable Metallurgy (IMPRS SusMet).
We acknowledge computational resources from the research center ZGH at Ruhr-University Bochum.
E.I acknowledges discussion with Christoph Freysoldt.

\bibliography{ace_water.bbl}

\end{document}